\documentclass[twocolumn]{aastex631}
\usepackage{etoolbox} 

\usepackage{amsmath, amsthm, amssymb}
\usepackage{wasysym}
\usepackage{enumerate}

\makeatletter
\let\original@label\label
\makeatother

\usepackage{hyperref}
\usepackage{cleveref}

\makeatletter
\let\label\original@label
\makeatother

\usepackage{multirow}
\shorttitle{JWST can detect Sub-Io-sized transits}
\shortauthors{Householder et al.}
\graphicspath{{./}{}}
\begin{document}
\title{Sensitivity to Sub-Io-sized Exosatellite Transits\\ in the MIRI LRS Lightcurve of the Nearest Substellar Worlds}

\correspondingauthor{Andrew Householder}
\email{andhouse@umich.edu}

\author[0009-0004-7103-7828]{Andrew Householder}
\affiliation{Department of Astronomy, University of Michigan, Ann Arbor, MI 48109, USA}

\author[0000-0002-9521-9798]{Mary Anne Limbach}
\affiliation{Department of Astronomy, University of Michigan, Ann Arbor, MI 48109, USA}

\author[0000-0003-4614-7035]{Beth Biller}
\affiliation{Scottish Universities Physics Alliance, Institute for Astronomy, University of Edinburgh, Royal Observatory, Blackford Hill, Edinburgh, EH9 3HJ, UK}
\affiliation{Centre for Exoplanet Science, University of Edinburgh, Edinburgh, EH9 3HJ, UK}

\author[0009-0008-5864-9415]{Brooke Kotten}
\affiliation{Department of Astronomy, University of Michigan, Ann Arbor, MI 48109, USA}
 
\author[0000-0003-3008-1975]{Mikayla J. Wilson}
\affiliation{Department of Astronomy \& Astrophysics, University of California, Santa Cruz, CA 95064, USA}

\author[0000-0003-0489-1528]{Johanna M. Vos}
\affiliation{School of Physics, Trinity College Dublin, The University of Dublin, Dublin 2, Ireland}
\affiliation{ Department of Astrophysics, American Museum of Natural History, Central Park West at 79th Street, NY 10024, USA}

\author[0000-0001-6098-3924]{Andrew Skemer}
\affiliation{Department of Astronomy \& Astrophysics, University of California, Santa Cruz, CA 95064, USA}

\author[0000-0001-7246-5438]{Andrew Vanderburg}
\affiliation{Department of Physics and Kavli Institute for Astrophysics and Space Research, Massachusetts Institute of Technology, Cambridge, MA
02139, USA}

\author[0000-0002-9962-132X]{Ben J. Sutlieff}
\affiliation{Scottish Universities Physics Alliance, Institute for Astronomy, University of Edinburgh, Royal Observatory, Blackford Hill, Edinburgh, EH9 3HJ, UK}
\affiliation{Centre for Exoplanet Science, University of Edinburgh, Edinburgh, EH9 3HJ, UK}


\author[0009-0005-9339-2369]{Xueqing Chen}
\affiliation{Scottish Universities Physics Alliance, Institute for Astronomy, University of Edinburgh, Royal Observatory, Blackford Hill, Edinburgh, EH9 3HJ, UK}
\affiliation{Centre for Exoplanet Science, University of Edinburgh, Edinburgh, EH9 3HJ, UK}

\author[0000-0002-1835-1891]{Ian J.\ M.\ Crossfield}
\affiliation{Department of Physics and Astronomy, University of Kansas, Lawrence, KS, USA}

\author[0000-0001-7866-8738]{Nicolas Crouzet}
\affiliation{Kapteyn Astronomical Institute, Rijksuniversiteit Groningen, Postbus 800, 9700 AV Groningen, The Netherlands}

\author[0000-0001-9823-1445]{Trent Dupuy}
\affiliation{Scottish Universities Physics Alliance, Institute for Astronomy, University of Edinburgh, Royal Observatory, Blackford Hill, Edinburgh, EH9 3HJ, UK}
\affiliation{Centre for Exoplanet Science, University of Edinburgh, Edinburgh, EH9 3HJ, UK}

\author[0000-0001-6251-0573]{Jacqueline Faherty}
\affiliation{ Department of Astrophysics, American Museum of Natural History, Central Park West at 79th Street, NY 10024, USA}

\author[0000-0001-7047-0874]{Pengyu Liu}
\affiliation{Leiden Observatory, Leiden University, PO Box 9513, 2300 RA Leiden, The Netherlands}
\affiliation{Scottish Universities Physics Alliance, Institute for Astronomy, University of Edinburgh, Royal Observatory, Blackford Hill, Edinburgh, EH9 3HJ, UK}
\affiliation{Centre for Exoplanet Science, University of Edinburgh, Edinburgh, EH9 3HJ, UK}

\author[0000-0003-0192-6887]{Elena Manjavacas}
\affiliation{AURA for the European Space Agency (ESA), ESA Office, Space Telescope Science Institute, 3700 San Martin Drive, Baltimore, MD, 21218 USA}

\author[0000-0003-2015-5029]{Allison McCarthy}
\affiliation{ Department of Astronomy \& The Institute for Astrophysical Research, Boston University, 725 Commonwealth Ave., Boston, MA 02215, USA}

\author[0000-0002-4404-0456]{Caroline V. Morley}
\affiliation{ Department of Astronomy, The University of Texas at Austin, Austin, TX 78712, USA}

\author[0000-0002-0638-8822]{Philip S. Muirhead}
\affiliation{ Department of Astronomy \& The Institute for Astrophysical Research, Boston University, 725 Commonwealth Ave., Boston, MA 02215, USA}

\author[0000-0001-5254-6740]{Natalia Oliveros-Gomez}
\affiliation{William H. Miller III Department of Physics and Astronomy, Johns Hopkins University, Baltimore, MD 21218, USA}

\author[0000-0002-2011-4924]{Genaro Su\'arez}
\affiliation{ Department of Astrophysics, American Museum of Natural History, Central Park West at 79th Street, NY 10024, USA}

\author[0000-0003-2278-6932]{Xianyu Tan}
\affiliation{Tsung-Dao Lee Institute, Shanghai Jiao Tong University, 1 Lisuo Road, Shanghai, 201210, People’s Republic of China}

\author[0000-0003-2969-6040]{Yifan Zhou}
\affiliation{Department of Astronomy, University of Virginia, 530 McCormick Rd., Charlottesville, VA 22904, USA}

\begin{abstract}
JWST's unprecedented sensitivity enables precise spectrophotometric monitoring of substellar worlds, revealing atmospheric variability driven by mechanisms operating across different pressure levels. This same precision now permits exceptionally sensitive searches for transiting exosatellites, small terrestrial companions to these worlds.
Using a novel simultaneous dual-band search method to address host variability, we present a search for transiting exosatellites in an 8-hour JWST/MIRI LRS lightcurve of the nearby ($2.0\,pc$) substellar binary WISE~J1049$\text{-}$5319\,AB, composed of two $\sim30\,M_{\rm Jup}$ brown dwarfs separated by $3.5\,au$ and viewed near edge-on. Although we detect no statistically significant transits, our injection-recovery tests demonstrate sensitivity to satellites as small as $0.275\,R_{\oplus}$ ($0.96\,R_{\rm Io}$ or $\sim$1 lunar radius), corresponding to 300\,ppm transit depths, and satellite-to-host mass ratios $>$$10^{-6}$. This approach paves the way for detecting Galilean-moon analogs around directly imaged brown dwarfs, free-floating planets, and wide-orbit exoplanets, dozens of which are already scheduled for JWST lightcurve monitoring. In our Solar System, each giant planet hosts on average 3.5 moons above this threshold, suggesting that JWST now probes a regime where such companions are expected to be abundant. The technique and sensitivities demonstrated here mark a critical step toward detecting exosatellites and ultimately enabling constraints on the occurrence rates of small terrestrial worlds orbiting $1\text{-}70$\,$M_{\rm Jup}$ hosts.

\end{abstract}


\section{Introduction} \label{sec:intro}

The transit technique continues to play a critical role in exoplanet detection. To date, it has been used to detect the vast majority of known exoplanets via ground and space-based surveys such as WASP \citep{2006PASP..118.1407P}, KELT \citep{Pepper2007}, CoRoT \citep{2009A&A...506..411A}, Kepler/K2 \citep{Borucki2010,Howell2014}, the HATNet project \citep{Bakos2018}, NGTS \citep{2018MNRAS.475.4476W}, and TESS \citep{Ricker2015}.

Transit surveys have generally been limited to visible wavelengths ($\lambda < 1\, \mathrm{\mu m}$), an ideal range for the detection of exoplanets orbiting main-sequence stars, including stars as cool as mid M-dwarfs \citep{2011Msngr.145....2J,2012AJ....144..145B}.
To expand the census of transiting companions to hosts of even lower masses, it is necessary to extend transit observations into the near-infrared \citep[NIR;][]{2019PASP..131k4401T,2021A&A...645A.100S}, where low-mass hosts are the brightest. Transit surveys operating beyond $\lambda=1$\,$\mu m$ can obtain higher precision lightcurves of late M-dwarfs, substellar worlds, and even directly imaged wide-orbit exoplanets as compared to studies like the MEarth project \citep{Irwin_2008} and SPECULOOS \citep{2021A&A...645A.100S}. Infrared transit searches can enable the detection of terrestrial transiting companions, which we refer to in this paper as {\it exosatellites}\footnote{Following a precedent established in the literature, we use the term ``exosatellite" or ``satellite" to refer to a companion orbiting substellar worlds \citep{2015ApJ...800..126K,2019BAAS...51c.169M,2022AJ....163..253T,2023PASP..135a4401L,2024AJ....168..175H,2024AJ....168...54L,Heller_2016,Limbach2021,2022MNRAS.516..391L}.}, around substellar worlds (M$\lesssim$75\,M$_{\rm Jup}$). 

Several attempts to conduct infrared transit surveys have been made previously. Ground-based NIR surveys \citep[e.g., PINES;][]{2019PASP..131k4401T,2022AJ....163..253T,2022AJ....164..252T} reach sub-Neptune mass companions, but cannot detect terrestrial satellites because of limitations due to the bright and variable sky background. A limited number of previous space-based IR transit surveys with Spitzer \citep{2017MNRAS.464.2687H,2018sptz.prop14131M,2019ESS.....430213M,2019sptz.prop14257M,2024AJ....168...54L} have demonstrated sensitivity to $\sim$Earth-sized satellites, but were limited in sensitivity by Spitzer's small aperture size as well as variability from the substellar worlds which can sometimes emulate transit signals. Such variability is considered the primary obstacle in searching for transits around substellar worlds.

With the advent of JWST, it is now possible to obtain { high precision} spectrophotometric lightcurves of brown dwarfs (\citealt{2024MNRAS.532.2207B}, GO 3496, 6474 and 8155), free-floating planets (\citealt{2025ApJ...981L..22M, 2024ApJ...977L..49R}, GO 3548 and 8846) and { wide-orbit} exoplanets (GO 3181, 3375, 4758, 5226 and 6139). The precision (reaching 10's of ppm on the most favorable targets) and small host sizes (all about the radius of Jupiter) presents an unprecedented opportunity to detect transits of exosatellites. The predicted sensitivity of these observations suggests that in many systems it will be possible to detect exosatellites smaller than the Galilean moons, and to reach satellite-to-host mass ratios as small as 10$^{-7}$ \citep{Limbach2021}. For comparison, there is an average of 4.5 moons per giant planet in our solar system that are of mass ratios $>10^{-7}$, and 3.5 moons per giant planet in our solar system that are of mass ratios $>10^{-6}$. Moreover, theoretical formation modeling predicts a similarly high occurrence rate of short-orbit exosatellites
\citep{Canup2006,cilibrasi2020nbody,2020A&A...638A..88L,2020MNRAS.499.1023I,2021MNRAS.504.5455C,2025NatCo..16.4853R}. Furthermore, the novel multi-band transit search technique demonstrated in this work presents a viable strategy for neutralizing the effect of host variability on transit searches. This in turn removes many concerns related to false-alarm detections induced by said variability, and dramatically increases the effect of satellite searches.

Here we demonstrate that JWST is capable of reaching this predicted sensitivity to extremely small transiting satellites using Cycle 2 observations of a pair of nearby substellar worlds. The layout of this paper is as follows, Section~\ref{sec:observations} covers details related to the dataset, how it was collected and the original motivations for doing so. Section~\ref{sec:analysis} derives the expected parameter values for our target exosatellite population. Following this, we detail our methodology and the results of our search. We outline our injection/recovery test procedure as motivated by our search result and list the results of our injection/recovery tests in Section~\ref{sec: I/R Results}. Finally, in Section~\ref{sec:discussion}  we discuss the wider implications of our study on the future of this field, and draw our conclusions in Section~\ref{sec:conclusion}.
\section{Observations} \label{sec:observations}

In this manuscript, we leverage JWST Mid-Infrared Instrument, Low Resolution Spectrometer (MIRI-LRS) data of the substellar binary objects WISE~J104915.57\text{-}531906.1\,A (hereafter WISE~J1049\,A), an L7.5 dwarf, and WISE~J104915.57\text{-}531906.1\,B (hereafter WISE~J1049\,B), a T0.5 dwarf \citep{2013ApJ...772..129B}. These objects have masses of $35.266_{-0.068}^{+0.067}\,M_{\rm Jup}$ and $29.476_{-0.057}^{+0.056}\,M_{\rm Jup}$ respectively \citep{2024AN....34530158B}. At $2.0\,pc$, WISE~J1049\,AB is the third closest system and the closest brown dwarf system to Earth. \cite{2023ApJ...945..119G} find that WISE~J1049\,AB is a member of the Oceanus moving group, constraining its age to $510\pm95\,Myr$. A full list of the substellar and system parameters is given in Table \ref{tab:parameters}.

The MIRI LRS observations utilized in this paper were conducted at UTC 12:24:19-21:05:33, on 8 July, 2023 as part of program GO 2965 PI Biller \citep[][hereafter B24]{2024MNRAS.532.2207B}. The observation was an 8\,hr time series observation (TSO) conducted with the LRS slitless mode using the P750L disperser. The traces of the two components were marginally resolved on the detector. B24 reported initial results from the program, including lightcurves of both A and B components. The data were processed as described in B24. However, for this analysis, we use the blended A+B lightcurve, as the individual lightcurves suffer from systematics that arrive from the close proximity of the two spectral traces on the detector.

\begin{table}[h]
\centering
\footnotesize
\caption{Summary of WISE~J1049\,AB Parameters}
\label{tab:parameters}
\setlength{\arrayrulewidth}{0.1mm}
\begin{tabular}{c|c|c|c}
Parameter & WISE~J1049\,A & WISE~J1049\,B & Ref \\ 
\hline
Name & \multicolumn{2}{c|}{WISE~J1049\text{-}5319\,AB} & 1 \\
Alt. Name & \multicolumn{2}{c|}{WISE~J1049\,AB} &  1 \\
Alt. Name & \multicolumn{2}{c|}{Luhman~16\,AB} & 2 \\
RA$_{\rm ICRS}$  (ep=2016.02) & \multicolumn{2}{c|}{$162.3282162^\circ\pm 6.1\,mas$} & 3\\
DEC$_{\rm ICRS}$  (ep=2016.02) & \multicolumn{2}{c|}{-$53.3194099^\circ$ $_{-6.1}^{+6.5}\,mas$ } & 3\\
RAcos(dec) PM [mas/yr] & \multicolumn{2}{c|} {-$2768.8551_{-0.0036}^{+0.0035}$} &3\\
DEC PM [mas/yr] & \multicolumn{2}{c|}{$358.4482\pm0.0055$} & 3\\
Orbital Separation [au] & \multicolumn{2}{c|}{$3.426\pm0.014$} & 3\\
Annual Parallax [mas] & \multicolumn{2}{c|}{$501.023_{-0.079}^{+0.077}$} & 3\\
Distance &\multicolumn{2}{c|}{$1.9960\,pc$$\pm50\,au$} & 3\\
${\rm T_{\rm eff}}$ [K] &\multicolumn{2}{c|}{$1150\text{-}1300$}&4\\
Radius [$R_{\rm Jup}]$ & \multicolumn{2}{c|}{1.00} & 4 \\
Age [Myr] & \multicolumn{2}{c|}{$510\pm 95$} &5\\
Mass [$M_{\rm Jup}$] & $35.266_{-0.068}^{+0.067}$ & $29.476_{-0.057}^{+0.056}$ & 3 \\
Spectral type & L7.5 & T0.5 & 6\\
Axial Tilt [deg] $^{a}$& $62\text{-}90$ & $87\text{-}90$ &7\\
\end{tabular}
\
\vspace{1mm}
$^{a}$We treat axial tilt as $70^\circ$ and $88.5^\circ$ respectively\\
{\footnotesize Refs: 1. \cite{2013ApJ...767L...1L}, 2.\cite{mamajek2013nearbybinarybrowndwarf}, 3. \cite{2024AN....34530158B}, 4. B24, 5. \cite{2023ApJ...945..119G}, 6. \cite{2013ApJ...772..129B}, 7. \cite{2021ApJ...906...64A}} 
\vspace{1mm}
\end{table}

\begin{figure*}
\centering
\includegraphics[width=0.9\textwidth]{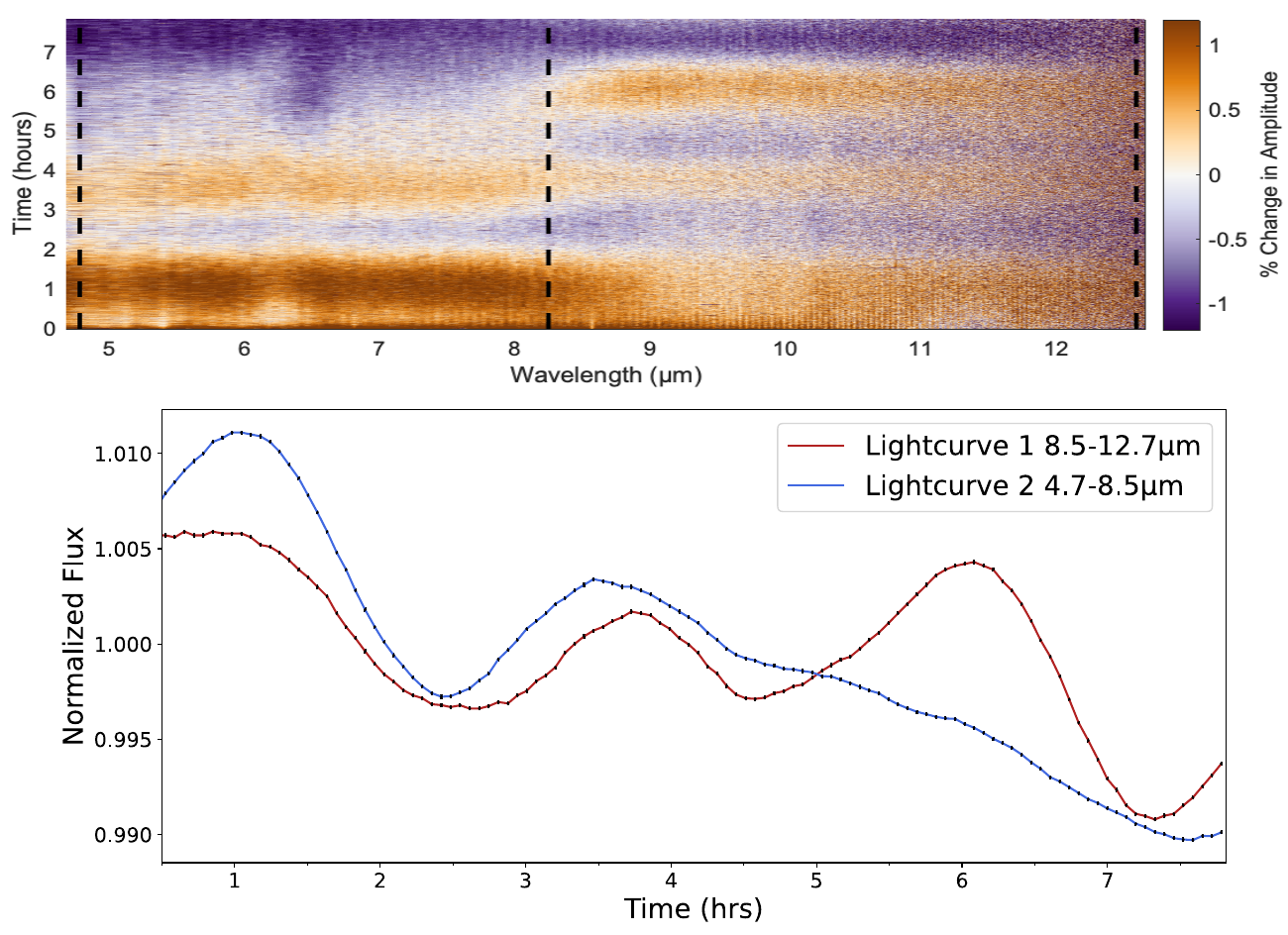}
\caption{{\bf Top:} Blended MIRI LRS variability map of WISE~J1049\,AB: wavelength (x-axis) vs. time (y-axis) and normalized flux (amplitude; colorbar). There are two distinct lightcurve behaviors present, allowing for the construction of two unique lightcurves distinguished by the change in behavior at $8.5\,\mu m$ (center dashed line). {\bf Bottom:} Lightcurves 1 (red) and 2 (blue) constructed from the above data and outlined with an $\sim11.77$\,min running median.}
\label{LCVarmap}
\end{figure*}

The primary goal of Program \#2965 was to study the weather and atmospheric drivers of variability of both substellar worlds, as described in B24. Alternatively, the exquisite sensitivity and spectrophotometric precision achieved with MIRI LRS on these substellar worlds is sufficient to enable a search for transiting exosatellites within the lightcurves. In this manuscript, we leverage the sum of the A and B lightcurves presented in B24 to conduct a search for small transiting rocky satellites. The data were processed as described in B24.

This program also collected timeseries observations with NIRSpec; however, the brightness of WISE~J1049\,AB required the use { of} only two groups per integration, rather than the recommended $\ge5$.  
As a result, the lightcurve exhibits a systematic signal well fit by a 6.5\,min sine wave with an amplitude of 0.2\,\%, possibly produced by heater cycling within the instrument.  
While this systematic does not hinder the program's primary objective (characterizing
substellar variability) it does limit our sensitivity to small transiting satellites.
Accordingly, we confine our transit search to the LRS data alone.

The blended (A+B) variability map is shown in Figure \ref{LCVarmap}. There is significant intrinsic flux variability present in the data due to inhomogeneous atmospheric features. Notably, there exists a change in variability structure at $8.5\,\mu m$, meaning two distinct lightcurves can be constructed from these data: lightcurve 1 corresponding to the $8.5\text{-}12.7\,\mu m$ region, and lightcurve 2 to the $4.7\text{-}8.5\,\mu m$ region. For each region, we computed the normalized flux-defined as the amplitude or percent deviation from the mean flux—across all wavelengths, as shown in the top panel of Figure \ref{LCVarmap}. The resulting lightcurves were then generated by taking the median amplitude at each time step, computed across all wavelengths within the respective spectral region. 

Variability features in the lightcurves of substellar worlds are capable of mimicking transit signals \citep{2024AJ....168...54L}. However, atmospheric variability is highly chromatic \citep{2024MNRAS.532.2207B,2025ApJ...981L..22M}, whereas transits are generally achromatic { (see discussion in section \ref{sec:5.3})}. The JWST MIRI lightcurves capture two distinct chromatic variability features. We have employed a novel strategy to use this color information to differentiate between signals that appear transit-like but vary with wavelength, which must be attributed to substellar variability, and signals that are both transit-like and wavelength invariant, which may be generated by a transiting exosatellite. 

\section{Analysis} \label{sec:analysis}

\subsection{Expected Transit Parameters}
Before conducting a search for transiting exosatellites, we quantify the expected transit durations, depths and transit probabilities for this system in order to better inform the bounds on our priors. We calculate the range of expected transit durations using the equation given by \cite{2010exop.book...55W}:
\begin{equation}\label{tdurEQ}
    t_{\rm dur} =\frac{P}{\pi}\sin^{-1}{\left[\frac{R_h}{a}\frac{\sqrt{(1+k)^2-b^2}}{\sin{i}}\right]},
\end{equation}
wherein $P$ is the orbital period, $R_h$ is the radius of the host, $a$ is the semi-major axis/orbital radius, $k$ is the ratio of satellite radius to host radius, $b$ is the impact parameter of the orbit, and $i$ is its inclination. Note that for our purposes, both `edge-on' and `equator-on' are used interchangeably to refer to an inclination of $90^\circ$. We assume that 1) the radii of WISE~J1049\,A and B are each ${1.00\,R_{\mathrm{Jup}}}$ following results obtained by B24; 
2) the orbit of any hypothetical exosatellite are circular, consistent with expectations for short-period satellites either due to formation conditions or subsequent dynamical evolution.

\begin{figure}
\centering
\includegraphics[width=0.463\textwidth]{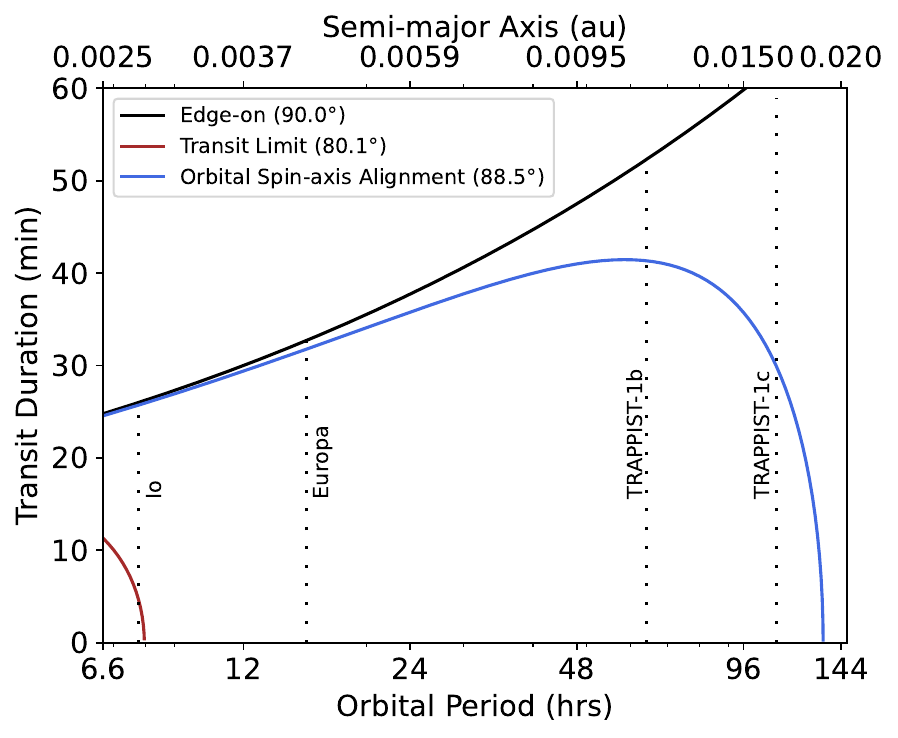}
\caption{Transit durations calculated for an Io-sized ($0.29R_{\oplus}$) exosatellite across multiple inclination cases. The x-axis begins at the Roche Limit  ($0.0025\,au$/$6.6\,hours$)  {and} extends to $144\,hour$ orbits at $0.02\,au$. An edge-on/equator-on ($90.0^\circ$, solid black line) case produces transits up to one hour. The case of spin-orbit alignment with WISE~J1049\,B ($88.5^\circ$, blue line) peaks at roughly 40 minute transits. The limiting case ($80.1^\circ$, brown line) shows the minimum possible inclination for which transits may still occur. The dotted vertical lines, from left to right, are the orbital periods/semi-major axes of Io ($0.0028\,au$/$7.8 hours$), Europa ($0.004\,au$/$15.9hours$), TRAPPIST~1b ($0.01154\,au$/$64.9\,hours$), and TRAPPIST~1c ($0.01580\,au$/$103.9\,hours$) \citep{2021PSJ.....2....1A}. These results only apply to the B component of the WISE~J1049\,AB system, but the small difference in mass between the two implies we would find similar results for the A component.}
\label{Tdur}
\end{figure}

\begin{figure}
\centering
\includegraphics[width=0.463\textwidth]{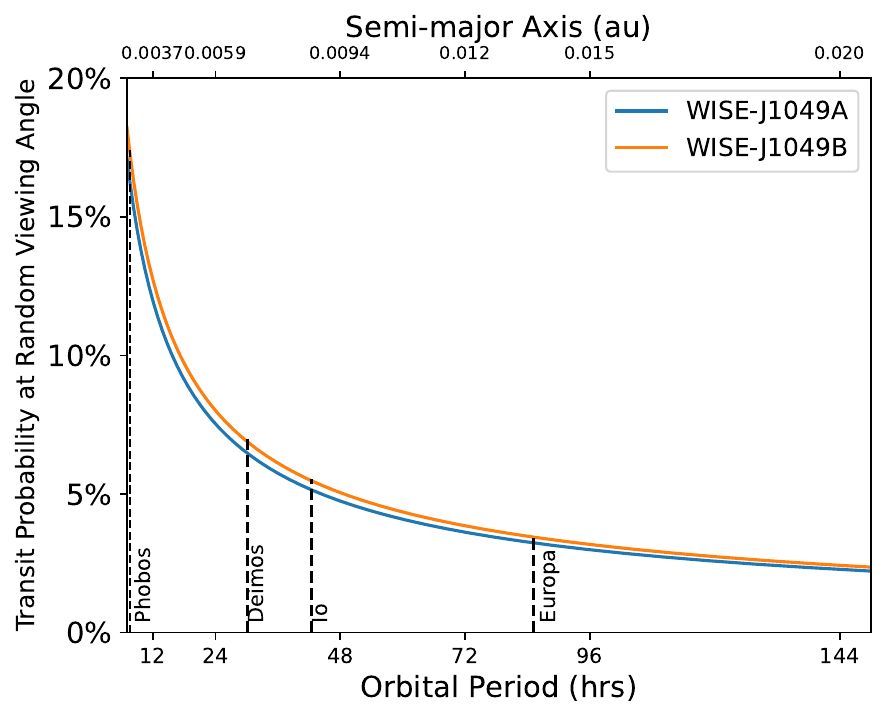}
\caption{The geometric probability of detecting a transit from a random viewing angle around either WISE~J1049\,A or WISE~J1049\,B individually as a function of the orbital period of the satellite. Reference lines are given at the orbital periods of Phobos ($7.6\,hr$), Deimos ($30\,hr$), Io ($42.5\,hr$), and Europa ($85.2\,hr$).}
\label{Tprob}
\end{figure}

Transit durations are dependent on the viewing angle inclination of the satellite system. In Figure~\ref{Tdur} we plot transit durations as a function of the semi-major axis of an orbiting Io-sized exosatellite in multiple inclination cases starting at the Roche limit and extending to $0.02\,au$ (about the semi-major axis of Trappist-1\,d or 3$\times$ the orbital separation of Jupiter's moon Ganymede). The Roche limit places an inner limit on any orbiting body, as objects orbiting within their Roche limit would be destroyed by tidal forces, the outer limit used in this plot is determined by our transit probability calculations (Figure~\ref{Tprob}) which show a negligible transit probability ($<2.5\%$) beyond this separation. Representing the longest possible durations is the edge-on inclination case (solid black line). Previous measurements constrain the spin-axis inclinations of WISE~J1049\,AB to $62\text{-}90^\circ$ and $87\text{-}90^\circ$ respectively, see Table \ref{tab:parameters} \citep{2021ApJ...906...64A}. Assuming that spin-orbit alignment could be preferential \citep{2023A&A...674A.120A}, we also plot the transit duration for a range of satellite viewing-angles corresponding to the spin-axis of the B component, specifically $88.5^\circ$ (blue line). For potential Io-radius exosatellites in edge-on alignment ($i=90^\circ$) with the B component, we calculate a maximum possible transit duration of $<1\,hour$. This value will set the upper limit on the transit duration prior when we conduct transit searches in this system. At the Roche limit of WISE~J1049\,B, transits should be visible for any exosatellites with inclination greater than $\sim{80.1^\circ}$ as indicated by the brown line. The spin-axis of A is constrained to a wider range of $62\text{-}90^\circ$. Under the assumption of spin-orbit alignment, transits would not occur for values $<80^\circ$. However, even a slight misalignment of the spin-orbit alignment by a few degrees in the right direction could allow transits around the A component, making it entirely reasonable for transits to occur around either host, even if the spin-axis of A is $<80^\circ$.

\begin{figure}
\centering
\includegraphics[width=0.463\textwidth]{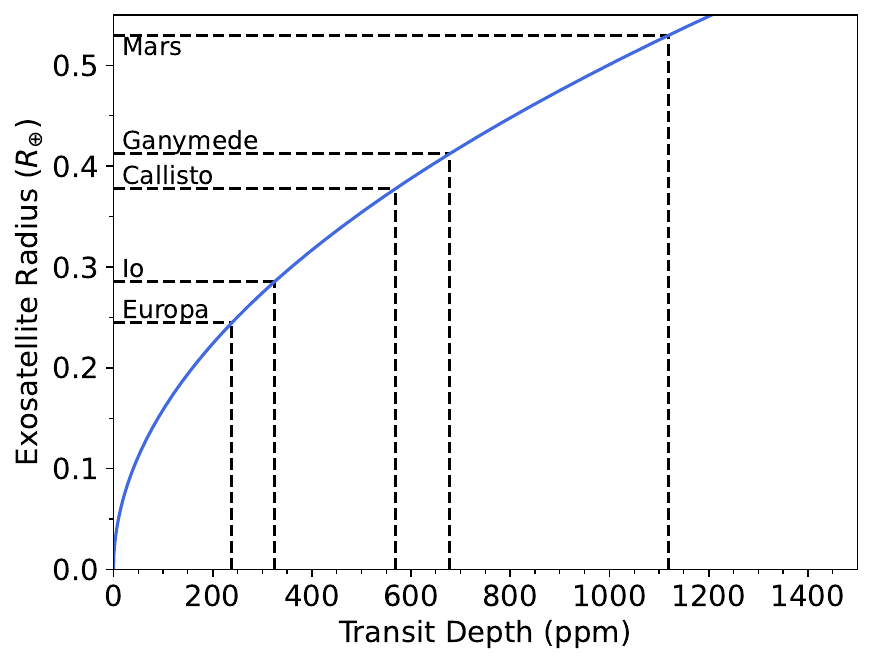}
\caption{The transit depths of the Galilean moons and the planet Mars as they would appear in orbit of WISE~J1049\,AB. Note that the transit depths computed here are for the unresolved A and B components, and would be $\sqrt{2}$ times larger if the lightcurves of the two components were resolved.}
\label{Tdepth}
\end{figure}

Since this study probes companion sizes smaller than those typically targeted in transit studies we provide context in the form of Figure~\ref{Tdepth}, showing possible exosatellite radii and their transit depths compared to Solar System objects. Our calculations reveal that analogues in radius to the Galilean moons of Jupiter would produce transit depths $200\text{-}700\,ppm$, whereas a Mars analogue would create a depth of $\sim{1100}\,ppm$, and an Earth analogue (not shown) would appear at $\sim{4000}\,ppm$. Because the lightcurves of the two substellar { objects} are unresolved, our transit depth calculations must be adjusted by a factor of $\sqrt{2}$ into the form:
\begin{equation}
    t_{depth} = \left(\frac{R_{s}}{\sqrt{2}~R_{h}}\right)^2
\end{equation}
where $R_{s}$ is the radius of the transiting exosatellite, and $R_{h}$ is the radius of the host body.
It is also true that because we keep the A and B components of the lightcurve blended together, we are unable to discern which of the binary objects is transited in the event of a detection. Fortunately, due to the approximately equal mass, radius, and luminosity of WISE~J1049\,A and B \citep{2024MNRAS.532.2207B}, our interpretation of satellite parameters from transit information remains accurate to the first order.  

Finally, we compute the geometric transit probability by dividing the host radius by the semi-major axis of the satellite, and plot as a function of orbital period, shown in Figure \ref{Tprob}. Assuming random viewing angle, probabilities range from $\sim3\text{-}20\%$. However, if we assume spin-orbit alignment in the B component ($i = 88.5^\circ$), transit probability remains at 100\% out to $\sim{0.02\,au}$ (6.1 day period). This implies that any satellite within this separation range is guaranteed to transit, provided its full orbital period is covered by the lightcurve observations.

\subsection{The Transit Search} \label{sec:TransitSearch}

\subsubsection{Methodology}
\label{sec:methods}
Now that we have quantified the parameter space of exosatellite transits expected in the WISE~J1049\,AB lightcurves, we search the JWST MIRI lightcurves for exosatellite transits that occurred during the observation. In this section, we will describe the search approach we are using and discuss our choice of priors.

We utilize custom built code to fit two models to the lightcurves:~1) a Gaussian process (GP) coupled with a transit model (GP+transit), and 2) a GP only (GPO) model. The GP model uses six parameters to fit each lightcurve (12 parameters in total) and is described in detail in \cite{2024AJ....168...54L}. The 6-parameter GP models are allowed to vary between the two lightcurves, meaning both models can account for chromatic variability of the hosts, while the 4-parameter transit model (described in the next paragraph) in the GP+transit fit is required to be the same in both lightcurves as the transit signal should be nearly achromatic.  
\begin{figure}
\centering
\includegraphics[width=0.475\textwidth]{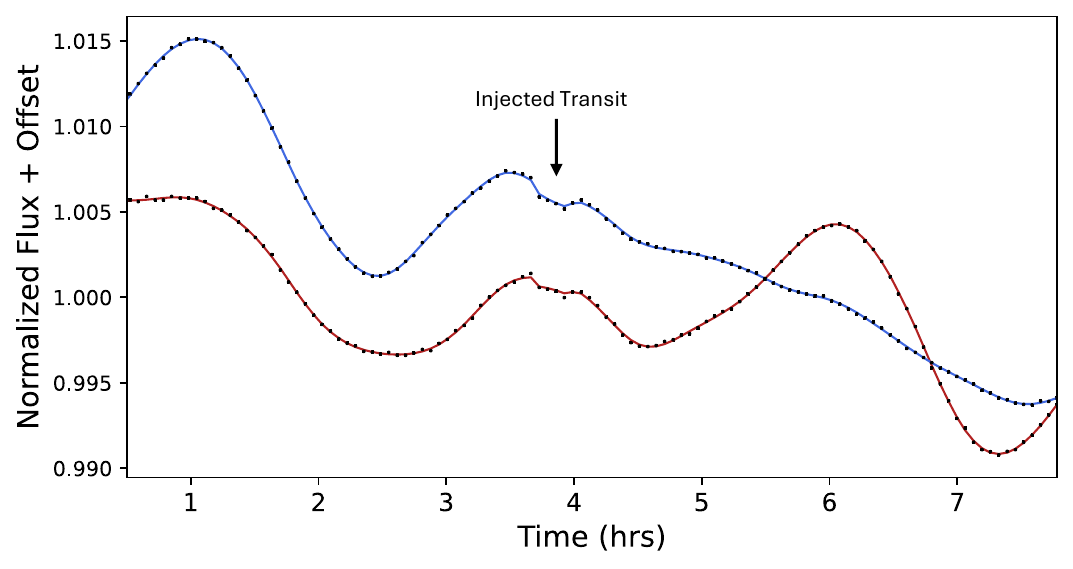}
\caption{WISE~J1049 data with an injection of a Mars-sized ($0.53R_\oplus$; $\sim1100\,ppm$) exosatellite, shown in both lightcurves 1 and 2 (black points). Red (Lightcurve 1) and blue (Lightcurve 2) curves show the GP+transit fits to the lightcurves, with the transit centered near {$4\,hours$} accurately captured by the transit model. Note that this transit is easily detected by eye despite the intrinsic variability of WISE~J1049\,AB, which has a larger amplitude than the transit itself.}
\label{Mars injection example}
\end{figure}

Because we are searching for transits near the sensitivity limit, we find that only a simple trapezoidal transit model is needed (e.g., our transit fits show no preference between transit models with and without limb darkening, therefore limb darkening is excluded). Our transit model is a trapezoid created using four transit-defining parameters: transit depth, the slope of ingress/egress, transit duration, and mid-transit time. When conducting our transit search we place priors on these parameters based on the theoretical calculations we made to characterize the range of exosatellite parameters we expect for this system. As a result, it would be unreasonable to expect transits to be detectable outside of this parameter space.

We adopt a uniform prior from zero to one hour for transit duration based on the results we derived in Figure \ref{Tdur}. We ensure that transit mid-times always remain within the bounds of the lightcurve, and we exclude the first 30 {minutes} of the lightcurves in our mid-time prior to allow for detector settling \citep{2023PASP..135c8002B,2024A&A...683A.212D}. Our priors place no restrictions on the impact parameter, with a uniform prior from zero to one. Large ($\geq$Mars-sized) transit depths are easily detectable by eye; for example, see Figure~\ref{Mars injection example}. Therefore, we focus our search on detecting only very small exosatellites and adopt a uniform prior ranging from 0 to 500ppm as larger transits would be seen by eye. However, an exception to this approach is discussed in Section~\ref{Injection Recovery}.

We fit our two models with the Python package \texttt{Dynesty} \citep{2020MNRAS.493.3132S}, which employs dynamic nested sampling. For the 16-parameter GP+transit model, we run the sampler with 1000 live points using the default bounding and sampling methods, to compute the evidence for each model, ${\rm log}z({\rm GP+T})$. For the GPO model, we run each lightcurve separately, as there is no need to perform this fit jointly since the transit fit (the only component that must be the same) is not included. When using a large number of live points to fit the six-parameter GPO model to the lightcurve, \texttt{Dynesty} fails to converge and times out. To address this, we run the sampler with only 60 live points, but perform three GPO runs for every one GP+transit run and in each of the two lightcurves. 
 From here we determine the fit parameters based on the median values from these three runs for each lightcurve and then computing the combined evidence ${\rm log}z({\rm GPO})$ for both lightcurves.
Additionally, we find that the ``\texttt{balls}'' bounding method in \texttt{Dynesty} is more efficient—achieving the best precision with minimal runtime—compared to the default bounding sampling methods. Therefore, we adopt this method for the GPO runs. 
For all subsequent analyses, we will report the resulting precision of the computed Bayesian evidence ($\Delta {\rm log}z$) for one model in comparison with the other.

After running \texttt{Dynesty}, we calculate the Bayesian evidence for both models, $\Delta {\rm log}z$, given by
\begin{equation}
    \Delta {\rm log}z = {\rm log}z\left({\rm GP+transit}\right)-{\rm log}z\left({\rm GPO}\right),
\end{equation}
to determine if there is statistically significant evidence favoring the GP+transit model ($\Delta {\rm log}z > 0$), or the GP-only model ($\Delta {\rm log}z < 0$). If the GP+transit model is favored, it suggests the presence of a transit-like event, potentially corresponding to an exosatellite transit in the lightcurve. For our purposes, we define a possible detection to be any case for which the $\Delta {\rm log}z \ge 1.55$, indicating marginal evidence favoring the GP+transit model over the GPO model \citep{Kass01061995}.

\subsubsection{Results of Transit Search}\label{ResultsTsearch}

Using this framework, we searched for transits in the WISE~J1049\,AB lightcurves. This resulted in a $\Delta {\rm log}z$ of $0.32\pm0.2$, indicating that evidence slightly favors the GP+transit model. However, this value remains well below our detection threshold of 1.55, suggesting that the evidence is not statistically significant. We conclude that there are no statistically significant transit signals detected in the lightcurve. 

We note that variability can produce transit-like signals, which may explain the slight evidence favoring the GP+transit model. If this is true, conducting the same test on inverted versions of the lightcurves should produce a similar $\Delta {\rm log}z$ result if the variability noise is symmetrical. 
This test further provides a calibration of variability's ability to mimic transits and serves as a secondary check against false positives caused by transit-like variability.
We performed our search for transits again, this time using the inverted (flipped upside down) versions of the lightcurves. This resulted in a $\Delta {\rm log}z$ value of $0.16 \pm 0.2$, again showing very slight favor toward the GP+transit model. However, the evidence remains far from statistically significant. We conclude that variability signals in the lightcurve are likely capable of producing extremely subtle transit-like and inverted transit-like signals, but since these are not above our detection threshold they do not produce false positives in our search.


\subsection{Injection/Recovery Testing} \label{Injection Recovery}

\begin{figure*}
\centering
\includegraphics[width=0.9\textwidth]{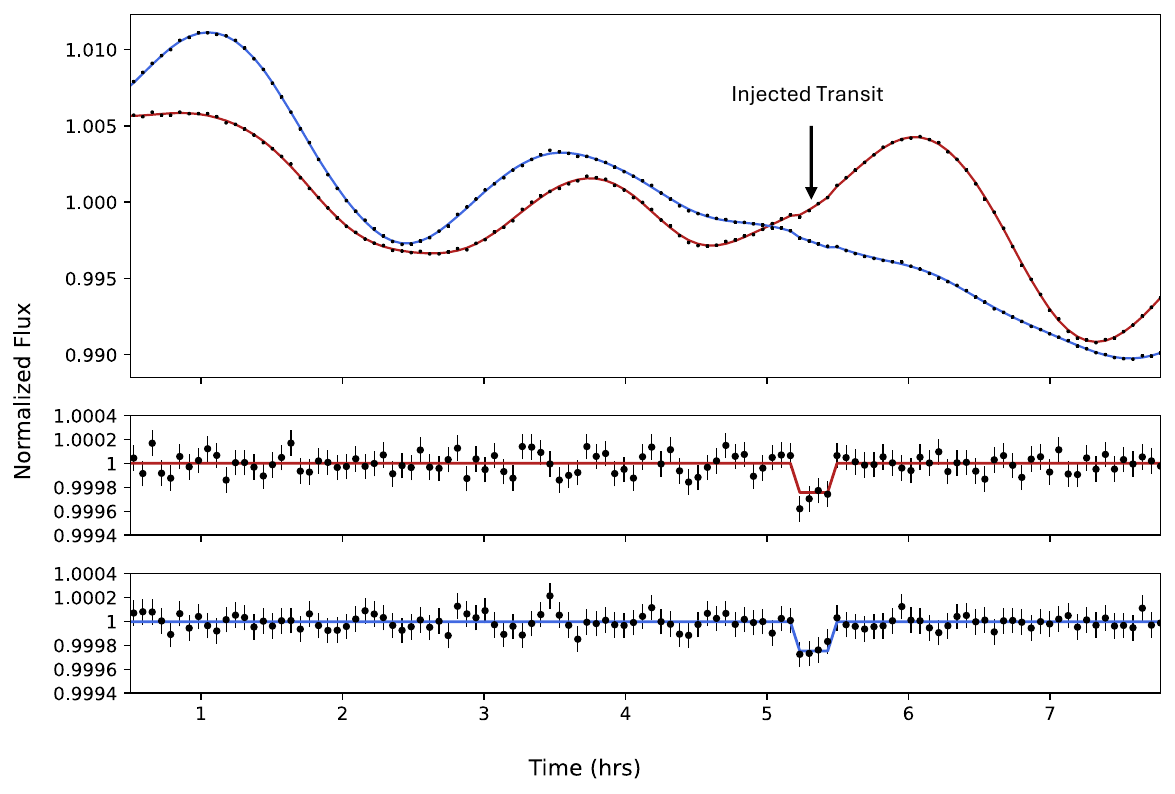}
\caption{Top Panel: WISE~J1049\,AB data (black points) with an injected $0.275R_\oplus$ ($0.96\,R_{\rm Io}$ or $\sim$1 lunar radius) exosatellite transit in both lightcurves 1 and 2 along with the GP+transit model fit (red and blue lines respectively). Bottom Panel: De-trended (from GP of the GP+transit fit) lightcurves highlighting the successful recovery of the injected transit with the GP+transit fit.}
\label{injection example}
\end{figure*}

\begin{figure}
\centering
\includegraphics[width=0.45\textwidth]{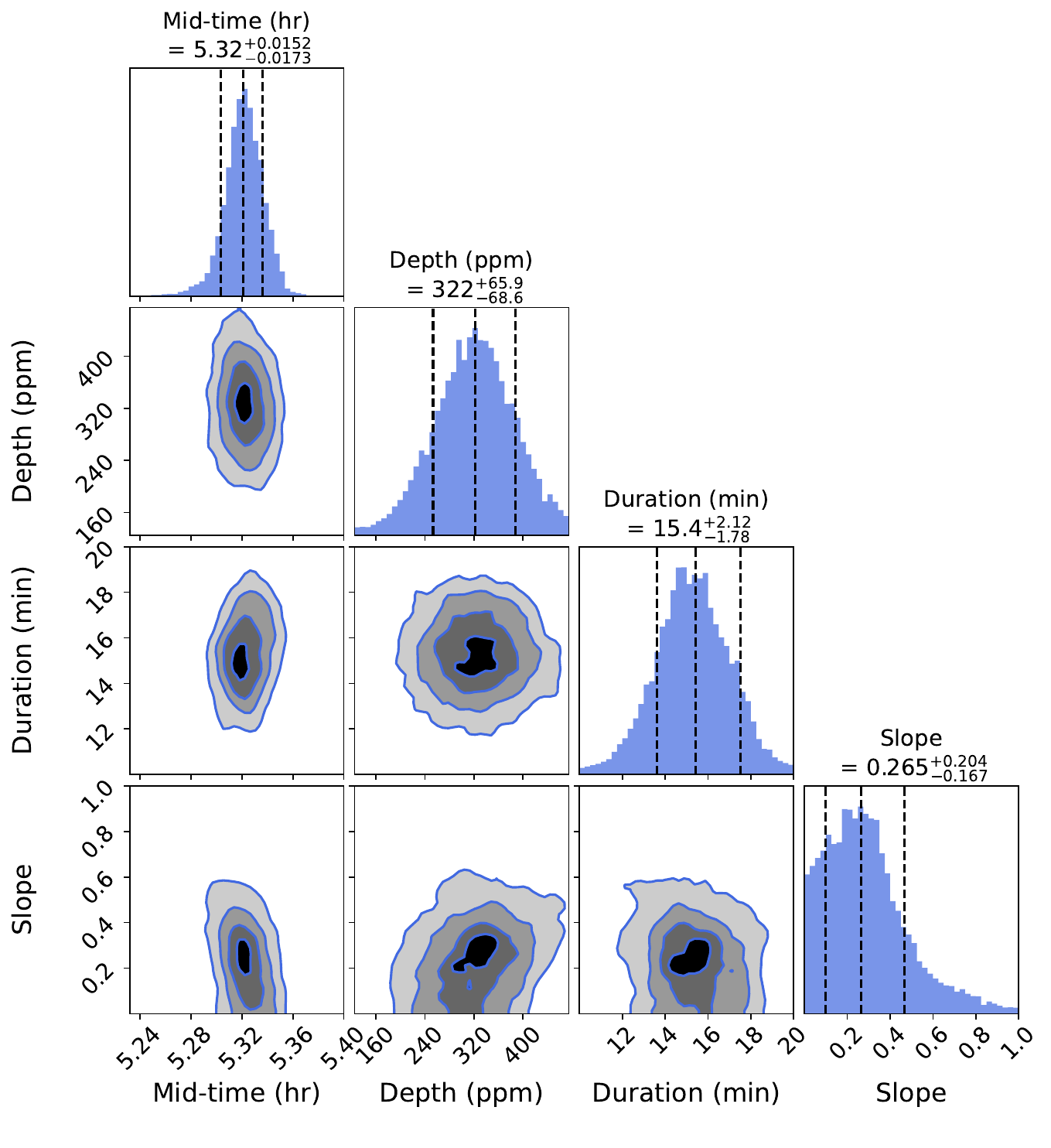}
\caption{Posterior parameter distributions for the transit model shown in Figure~\ref{injection example}. We see well constrained values on all transit defining parameters with the exception of the slope of ingress/egress. The injected transit depth was { 0.030\% ($\sim300\,ppm$)}, and the retrieved transit depth was { $0.032\pm0.007\%$.}}
\label{Example Injection Corner}
\end{figure}

Although no transits were detected in the WISE~J1049\,AB lightcurves, we conduct a series of transit injection and recovery tests to evaluate the range of parameter space of exosatellites transits that could have been detected around these substellar worlds had a transit occurred during observation. This exercise aims to quantify the parameter space where JWST is sensitive to transiting exosatellites. 

\begin{figure*}
\centering
\includegraphics[width=0.97\textwidth]{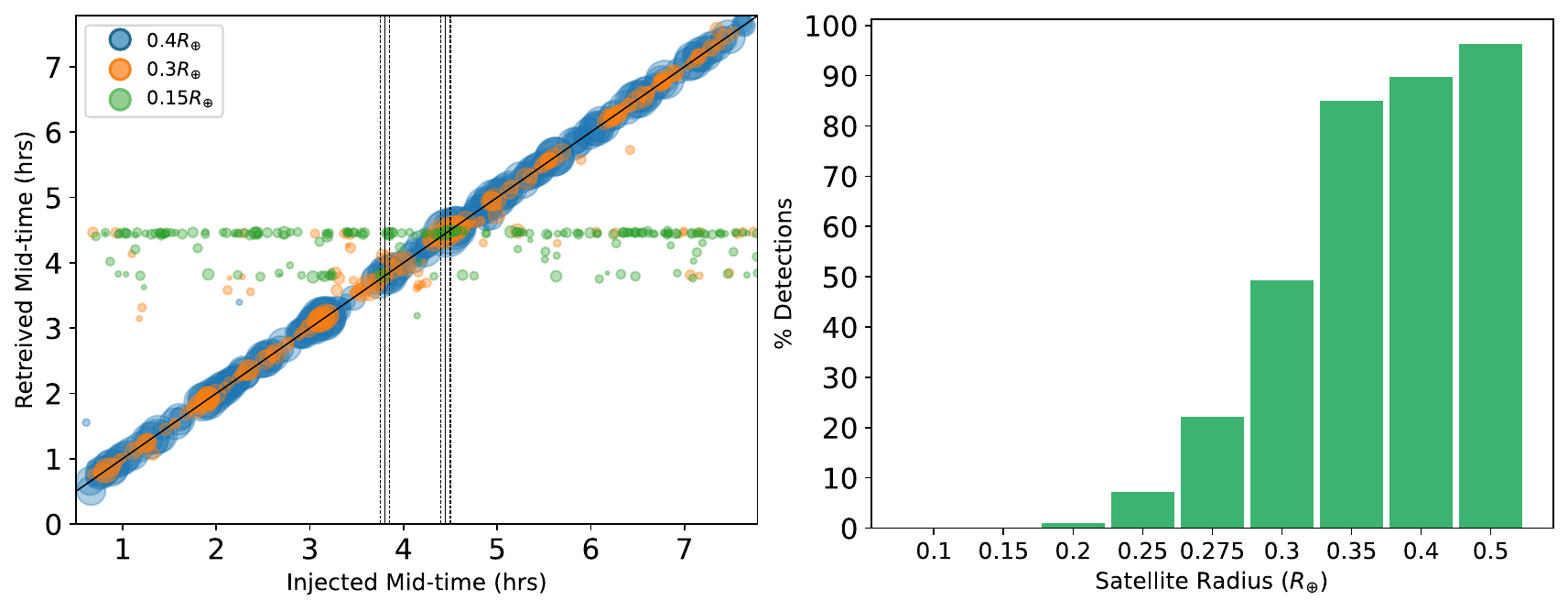}
\caption{Left Panel: A plot of the injected and recovered mid-transit times for three of our nine exosatellite radius cases. {The size of each point is scaled to the $\Delta {\rm log}z$ value of the test it represents, with the largest points having the largest $\Delta {\rm log}z$, and thus the highest favor of a transit inclusive model.} Two ``exclusion zones" corresponding to two variability features in the lightcurves which create transit-like signals (that are not statistically significant) at $t\sim3.8\,hours$ and $t\sim4.5\,hours$ are marked by vertical lines. All transits injected into these regions ($3\%$ of the total set) are excluded from injection/recovery statistics. Right Panel: A bar chart showing the percentage of recoveries made as a function of satellite radius. A transit is considered successfully recovered if it meets two metrics: 1) a recovered mid-time of transit within 3\,min of the injected value and 2) a $\Delta {\rm log}z>1.55$.}
\label{recovery_barchart}
\end{figure*}

To conduct this test, we inject transits into the two lightcurves over a wide range of transit depths, orbital separations, and inclinations. When injecting transits, we randomly select orbital separations from the range $0.002\,au < a < 0.06\,au$. The lower bound on orbital separation is given by the Roche limit of WISE~J1049\,B and the upper bound, 0.06\,au is slightly beyond the point at which transit probability drops to $<$1\% (assuming a random viewing angle) and the orbital period is $\sim$20 days, 60$\times$ longer than the observation duration (Figure \ref{Tprob}), meaning the probability of detecting satellites at larger orbital separations becomes negligible. For each injection the inclination is randomly selected from a range of $87.0^\circ < i < 90^\circ$. Although transits can occur at smaller inclinations for the smallest orbital separations, we choose this range for injections as it corresponds to a spin-axis alignment case of the B component. The fourth transit-defining parameter, mid-transit time, is selected randomly from within the lightcurve duration, though it excludes the first 30 minutes as previously discussed. Randomly selecting these parameters for each injection ensures that we test detection capabilities across the full range of plausible transit space. An example injection is shown in Figure \ref{injection example}, which demonstrates a $0.275R_\oplus$ ($0.96\,R_{\rm Io}$ or $\sim$1 lunar radius) satellite injected into the WISE~J1049 lightcurves. The posterior parameter distributions corresponding to this example injection as found by our transit searching code are displayed in Figure \ref{Example Injection Corner}.

Rather than randomly selecting the transit depth of each injection, we step through set values of exosatellite radii ranging from 0.1-0.5$R_{\oplus}$ in increments of 0.05$R_{\oplus}$ (excepting 0.45$R_{\oplus}$ and including 0.275$R_\oplus$), corresponding to transit depths between $40\text{-}995\,ppm$. We did not create any models of exosatellites  larger than 0.5$R_{\oplus}$, as we found that above this satellite radius, we successfully recovered 100\% of all injections that were not grazing the brown dwarf's limb (as mentioned before, such large transits are also detectable by eye, e.g., Figure \ref{Mars injection example}). We initially conducted the injection/recovery test over a radius range of 0.1\text{-}0.35$R_{\oplus}$ but found that the detection rate for the largest transit depth, 0.35$R_{\oplus}$, was just below 100\%. Therefore, we expanded our injections up to a radius of 0.5$R_{\oplus}$. However, this required a broader prior on the transit depth, so we used a uniform prior of $0\text{-}1000\,ppm$.  For each exosatellite size case, we conduct 200 injection and recovery tests. Across the nine exosatellite radius cases, this totals 1800 simulated transits (about $\sim1000$\,hrs of computing time).

\begin{figure*}
\centering
\includegraphics[width=0.97\textwidth]{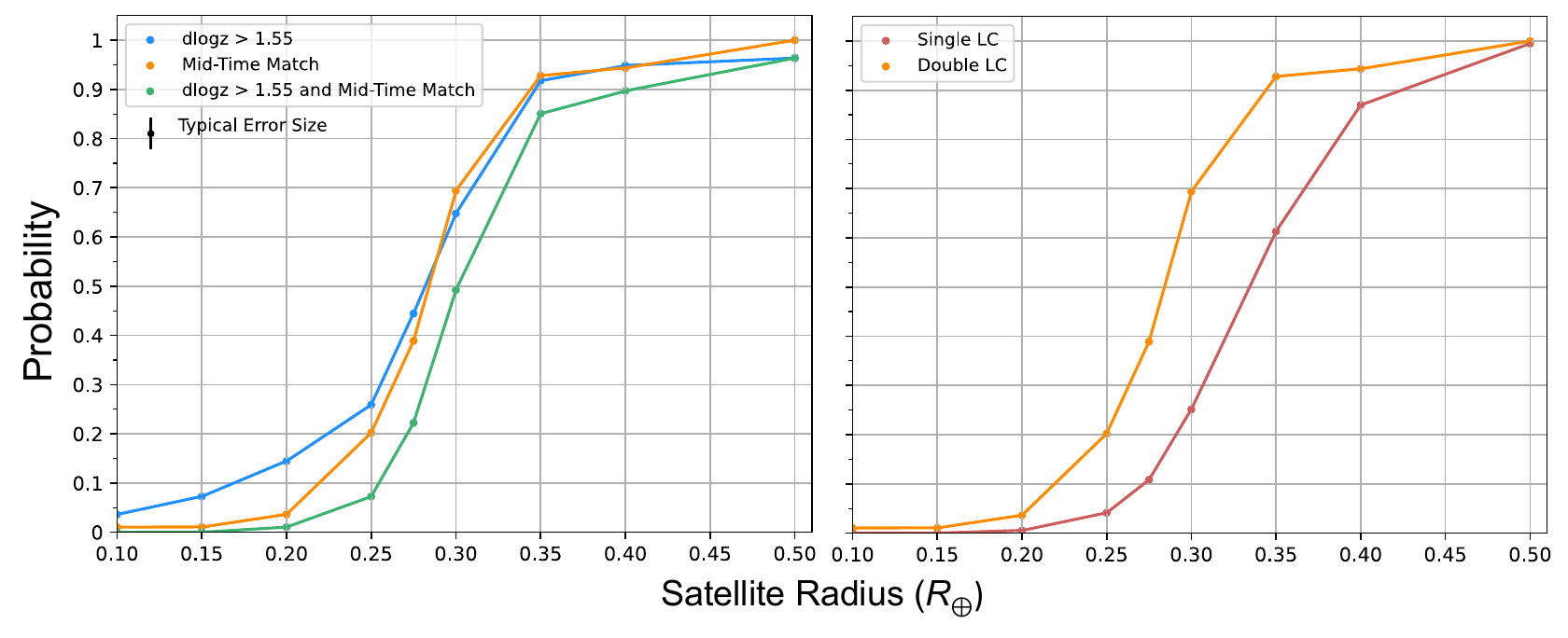}
\caption{{Left Panel:} The probability of retrieving a transit for any one injection/recovery test is tracked on the y-axis across the range of injected satellite sizes on the x-axis for the different recovery metrics. The detection rate reaches $20\text{-}50\%$ (depending on the metric) for injected exosatellites of radius $0.275R_\oplus$ (0.96\,R$_{\rm Io}$ or $\sim$1 Lunar radius). For satellites $>0.35R_\oplus$ (0.85\,R$_{\rm Ganymede}$), nearly all transits are detected. The typical error value is calculated by first separating the results of each individual size bin into two sub-bins containing half of the injected transits. The average difference between the number of injected transits which meet metric 1 falling into each sub-bin, for each size bin is multiplied by $\sqrt2$ to give the typical error. {Right Panel: A comparison between the probability of a transit injection/recovery test meeting metric 1 using our multi-band approach (orange curve, matching that in the left panel) versus using a single-band approach (red curve).}}
\label{falsepositiveplot}
\end{figure*}

\section{Results of Injection \& Recovery Testing} \label{sec: I/R Results}

After injecting a transit into the lightcurve, we follow the same procedure as in the injection-free search (Section~\ref{sec:methods}), utilizing both the GP+transit and GPO model fits to attempt recovery of the 1800 exosatellite transit signals. We use two metrics to assess whether a transit has been detected: 1) any transit for which the mid-transit time is recovered to within three minutes of the injected mid-transit time is considered to be a successful recovery, and 2) we consider the GP+transit model to be favored over the GPO model if it has a $\Delta {\rm log}z$ greater than 1.55, consistent with statistically significant evidence in favor of the GP+transit model \citep{Kass01061995}.

Figure \ref{recovery_barchart} shows the recovered vs. injected mid-time of the transits in three size cases (left panel), as well as the number of recoveries meeting both metrics 1) and 2) as a function of satellite radius (right panel). As shown in the left panel of this figure, there are two features near 3.8 and 4.4 hours where the GP+transit model becomes stuck and recovers transit signals (that are not statistically significant) as the transit depth becomes very small. When computing the number of detections as a function of satellite radius (right panel), we exclude all transits injected within 3 minutes of these points. This is because the retrieval process tends to get stuck in these regions due to an existing signal in the data, that causes detections regardless of whether the injected transit was actually identified. To avoid this ambiguity, we omit these cases from our analysis as including these cases would likely lead to an overly optimistic estimate of our recovery rate. Approximately $3.1\%$ of our injections fall within this time range, meaning that, on average, only about 194 out of the 200 injections in each size bin are used for our analysis.

The results of the {injection/recovery} testing are summarized in Figure \ref{falsepositiveplot} which shows three metrics corresponding to recovery percentages as a function of injected satellite radius { in the left panel}. As a function of injected radius, we report the values 1) and 2) described above, as well as 3) injected transits for which both of the previous criteria are true.

As injected satellite size increased, we recover an increasing percentage of injected transits, with the sharpest increase occurring between $0.25$ and $0.35\,R_{\oplus}$. The detection rate for injected exosatellites with a radius of $0.275\,R_\oplus$ ranges from $20\text{-}50\%$, depending on the metric used. For satellites with radii greater than $0.35\,R_\oplus$ ($\sim$85\% the radius of Jupiter's moon Ganymede), nearly all injected transits are recovered. { The right panel shows the effect of performing the same transit search with only a single lightcurve. In comparison, our multi-band search recovers 15\% more of the injected transits than with the single-band approach.}

\section{Discussion} \label{sec:discussion}

\subsection{A Path to Constraining Occurrence Rates of Small Terrestrial Exosatellites}

Our analysis demonstrates sensitivity to exosatellites in a parameter space where the occurrence rates are expected to be abundant. Theoretical predictions of exosatellite occurrence rates \citep{cilibrasi2020nbody,2020A&A...638A..88L,2020MNRAS.499.1023I,2021MNRAS.504.5455C} suggest that short-orbit exosatellites should be common around substellar worlds and wide-orbit exoplanets. The moons of the Solar System's giant planets indicate similarly high occurrence rates, with an average of 3.5 moons exceeding a satellite-to-host mass ratio of $10^{-6}$ (the sensitivity achieved in this study) with orbital periods between $0.9\text{-}17\,days$ \citep{2024AJ....168...54L}.

By the end of Cycle 3, JWST will have monitored the lightcurves of seven wide-orbit exoplanets (Beta~Pic\,b, HR 8799\,b,\,c,\,d\,\&\,e, 2M1207\,b, VHS~J1256\,b and Ross~458\,c) and eight free-floating substellar worlds (WISE~J0855\text{-}0714, WISE~J1049\,A\&B, SIMP~J0136+09, 2MASS~1047, 2M1237 and W1122), including the WISE~J1049\text{-}5319\,AB system. While these observations may not reach the same precision as the data used in the study, each of these observations offer an opportunity to search for transiting exosatellites. We also note that despite WISE~J1049\,AB's proximity and brightness, the results presented here are likely not the most optimistic scenario as: 1) we take a $\sqrt2$ hit on sensitivity because the lightcurve of the two substellar worlds is blended and 2) Observations of substellar worlds in the NIR (rather than mid-IR as presented here) generally achieve a higher precision (lower photon-noise limit), except on extremely cold worlds \citep[e.g., $\lesssim$300\,K;][]{2024Natur.633..789M,2025arXiv250515995B}. 

\begin{figure}[h!]
\centering
\includegraphics[width=0.45\textwidth]{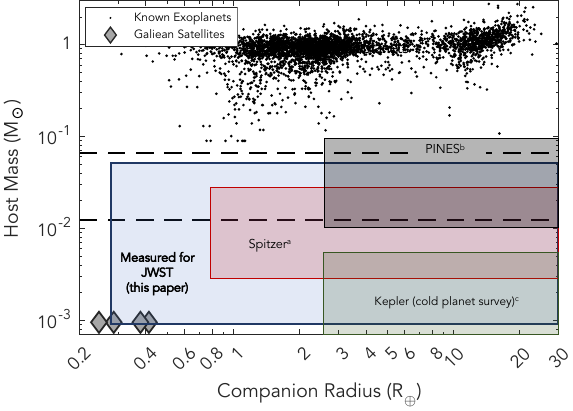}
\caption{Companion { radius} vs. host mass of the known exoplanet population ({ black points}; includes only confirmed planets with measured { radii}) as well as the approximate sensitivity of several past or ongoing exosatellite (or moon) surveys of substellar, FFP and exoplanet hosts. This study demonstrates JWST is sensitive to exosatellites down to $0.275R_{\oplus}$ around $30\,M_{\rm Jup}$ hosts. JWST is set to observe the lightcurves of dozens more wide-orbit exoplanets, FFPs, and brown dwarfs. If the survey's sensitivity applies to other host types (see discussion), the resulting JWST lightcurves will explore the exosatellite parameter space indicated by the light { blue} shaded region in this plot. 
 References: (a) \citealt{2024AJ....168...54L}, (b) \citealt{2022AJ....163..253T,2022AJ....164..252T} , (c) \citealt{2022NatAs...6..367K}}
\label{Discussion Plot}
\end{figure}

Given the predicted transit probabilities for short-orbit satellites and their anticipated occurrence rates ($5\text{-}15\%$, depending on host mass, for a satellite with the same orbital period of Io; \citealt{Limbach2021}), our statistical estimates indicate that observing about a dozen such objects should result in the detection of an exosatellite, assuming that satellite-to-host mass ratios of at least $10^{-6}$ are detectable with the other datasets. Based on the current rate of observations, JWST is likely to capture dozens, if not hundreds, of substellar and directly-imaged exoplanet variability lightcurves during its lifetime. This data, coupled with the technique presented herein, provides an unprecedented opportunity to constrain the occurrence rates of exosatellites around $1\text{-}70\,M_{\rm Jup}$ objects. This host-mass range remains largely unexplored \citep{2022AJ....163..253T,2022AJ....164..252T,2022NatAs...6..367K,2024AJ....168...54L}, and probing it will significantly advance our understanding of satellite formation and dynamics. Any detected exosatellites or exomoons will likely span the gap between the moons of the Solar System giant planets and exoplanets of very low mass M-dwarfs, and provide our first constraints on occurrence rates of terrestrial companions within the substellar host-mass range (see Figure~\ref{Discussion Plot}). Figure~\ref{Discussion Plot} shows the companion and host masses of the currently known exoplanet population, in comparison to those of the Galilean moons of Jupiter. Four colored regions are shown indicated the range of transit sensitivity provided by previous exosatellite or moon surveys in green, yellow, and blue. The range of sensitivity determined by the results of this work is shown in the red colored field, which extends into a previously unexplored area of mass-ratio space by probing significantly lower satellite masses which approach those of Jupiter’s Galilean moons. This plot shows that JWST uniquely allows us to identify a new class of object that enables us to place the moons of our Solar System within the context of the broader population of extrasolar moons and satellites.

\subsection{Using Multi-Band Lightcurves to Disentangle Variability from Transits}

Generally, transit searches, including exoplanet transit searches, are done with single band observations (e.g., Kepler, TESS, HAT-NET, WASP). { Note that although CoRoT conducts multi-band monitoring \citep{2009A&A...506..287L}, it does not make use of the multiple bands for transit searching}. The use of dual- or multi-band monitoring to detect exoplanet transits has been discussed a few times in prior literature. The idea was first proposed by \citealt{1971Icar...14...71R} who suggest the theoretical idea that observing transits in two colors could help distinguish true planetary signatures from stellar or atmospheric noise, since a planet’s transit would dim both wavelengths equally, while stellar variability or Earth's atmosphere often has chromatic effects.

\citealt{2020AJ....160..240G} introduced a 2D Gaussian Process method that models correlated noise across time and wavelength using a shared kernel structure, ideal for multi-wavelength transit fitting and transmission spectroscopy. In contrast, our approach fits fully independent GPs to two lightcurves while enforcing a shared, achromatic transit model—explicitly leveraging the chromatic nature of variability to distinguish it from genuine transits. This makes our method better suited for detecting subtle exosatellite signals in the presence of time-variable host behavior. 

The multi-band monitoring and novel transit search approach presented here may be valuable for a broader range of transit search applications where host variability dominates over photon noise or other non-astrophysical noise sources. This includes young stars {\citep{2024Natur.635..574B, 2025AJ....169...27H,2017AJ....154..224R}}, which exhibit significant variability, as well as bright stars, where the dominant noise source can be stellar granulation {  and M dwarfs \citep{2024AJ....167..284G}}. In such cases, pushing to smaller planet radii would require accounting for granulation, which occurs on timescales comparable to planetary transits in K-type and G-type stars \citep{2017A&A...597A..94C}. {  Similar simultaneous fitting techniques have also been used to try disentangling periodic transit signals from systematic noise \citep{{2016A&A...585A..57K}}.} { Our simultaneous multi-band approach may also prove useful beyond the search for exosatellites, proposed surveys such as Argus \citep{2022PASP..134c5003L} will include a two-band optical system may take advantage of this technique in searches for exoplanets orbiting main-sequence stars.}

Another unique aspect of this work is our method for constructing the lightcurves. Unlike stellar granulation, which evolves slowly and coherently across wavelengths, variability in substellar objects tends to remain constant over broad spectral regions and then change abruptly, typically at wavelengths associated with absorption by specific atmospheric species or clouds \citep{2024MNRAS.532.2207B,2025ApJ...981L..22M,10.1093/mnras/staf737}. We exploit these spectral transitions (in the case of WISE~J1049, likely caused by silicate clouds in the mid-infrared) to construct lightcurves that are as orthogonal as possible, maximizing the differences in the variability behavior between them. This strategy may also be useful for transit searches in other contexts where abrupt changes in the variability pattern occur across different wavelengths. However, constructing such optimized lightcurves requires either access to spectrophotometric time-series data or instrumentation capable of multi-band observations with filters tuned to known variability features of the target.

{
\subsection{Are Exosatellite Transits Really Achromatic?}\label{sec:5.3}
In this study, we have assumed that the substellar variability signal is wavelength-dependent, while the exosatellite signal is not. However, this assumption may break down for extremely high precision transit measurements for several reasons. First, the atmospheres of companions can produce wavelength-dependent transit depths \citep{2023AJ....165...23T}. Terrestrial exosatellites, however, are expected to have very thin atmospheres: WISE~J1049\,AB has a size and brightness comparable to TRAPPIST-1, and thus TRAPPIST-1 provides a useful reference point. The atmospheres of the TRAPPIST-1 planets are generally undetectable with JWST unless 10's to 100's of transits are stacked together \citep{2019AJ....158...27L}. The satellites we are searching for are of similar size to the TRAPPIST-1 planets or smaller, closer to large Solar System moons such as Titan. Titan’s low surface gravity produces a more extended atmosphere with scale heights of 15–50 km, compared to 5–8 km on Earth \citep{2017JGRE..122..432H}. While this increases the atmosphere-to-solid-body ratio by up to an order of magnitude relative to Earth, the effect remains negligible for this observation. For example, a $\sim$50~km atmosphere atop Titan’s $\sim$2500~km radius would require a 4\% precision on the measured transit depth in the most favorable case for detection. However, for a Titan-sized planet, this observation would yield only $\sim$10\% precision in transit depth, rendering such an atmospheric signature below the $1\sigma$ detection threshold.

Second, it is known that heterogeneities in host star photospheres can produce chromatic exoplanet transit signals \citep{2018ApJ...853..122R,2019AJ....157...96R}. Similarly, substellar wolds may have zonal regions with and without clouds may introduce wavelength dependence for exosatellites. WISE~J1049\,AB, for instance, is known to have such cloud structures \citep{2014Natur.505..654C,2024MNRAS.533.3114C}, and most substellar objects exhibit similar features. While a detailed treatment is beyond the scope of this paper, we expect this to also be a secondary effect. As a reference point, Jupiter’s belts and zones differ in brightness by $\sim$20\% \citep{2019AJ....157...89G}, but the chromaticity of these variations is typically $\lesssim$10\% of the total effect. Thus, although an exosatellite transiting brighter or darker regions could slightly alter its inferred size, the chromatic contribution is likely only a few percent. We therefore conclude that treating the transits as achromatic remains a reasonable approximation for the precisions achieved in this study.  
}

\section{Conclusion} \label{sec:conclusion}

We have carried out a search for transiting exosatellites in the WISE~J1049$\text{-}$5319\,AB system, a nearby substellar binary composed of two $\sim$30\,$M_{\rm Jup}$ brown dwarfs with radii comparable to Jupiter.

\begin{itemize}
    \item Leveraging 8 hours of JWST MIRI-LRS spectroscopic monitoring, we searched for, but did not detect, any statistically significant transit events during the observation.
    \item To conduct this search, we developed a novel dual-band transit detection technique that simultaneously analyzes two lightcurves, each with an independent GP fit, while enforcing a common transit model. Since variability is typically chromatic and transits are approximately achromatic, this approach provides a powerful means of distinguishing variability-induced features from true transit signals. We discuss how this method may benefit other transit searches affected by stellar variability, such as those targeting young, active stars.
    \item This is the first transit search conducted in the mid-IR ($\lambda>5\,\mu m$), demonstrating that reaching high precisions (of a few hundred ppm) are possible and that long-wavelength transit searches are ideal for searching for exosatellites around substellar worlds. For comparison, previous near-IR observations of WISE~J1049$\text{-}$5319\,AB using the Hubble Space Telescope reached precisions between 2000-4000\,ppm \citep{Buenzli_2015}
    \item We performed transit injection–recovery testing to assess our sensitivity across parameter space. This exercise demonstrated the capability to detect transits as shallow as 300\,ppm, corresponding to a satellite radius of $0.275\,R_{\oplus}$ {($0.96\,R_{\rm Io}$ or $\sim$1 lunar radius)} and a satellite-to-host mass ratio of approximately $10^{-6}$.
    \item Noted that, in our Solar System, each giant planet hosts on average 3.5 moons above the aforementioned size threshold. This places our detection regime squarely in a parameter space where such companions are expected to be common, enabling the search for Galilean moon analogs orbiting directly imaged brown dwarfs, free-floating planets, and wide-orbit exoplanets (see Figure~\ref{Discussion Plot}), many of which are already slated for JWST monitoring.
 
\end{itemize}

The level of sensitivity achieved in this study demonstrates JWST’s capability to detect small transiting companions around substellar objects. For comparison, only one known transiting exoplanet smaller than $0.35\,R_{\oplus}$, Kepler-37\,b, has been discovered around a star, requiring 978 days of \textit{Kepler} observations and $\sim$73 transits. By contrast, our analysis shows that similarly sized satellites could be detected with a single transit in JWST data, underscoring the advantages of monitoring compact hosts. The techniques and sensitivities demonstrated here represent a path forward towards detecting exosatellites and ultimately constraining the occurrence rates of small, terrestrial companions around $1\text{-}70$\,M$_{\rm Jup}$ hosts.

\section*{Acknowledgments}

{We would like to thank our reviewers for their input, this paper has been made all the better for their insight.}

This work is based on observations made with the NASA/ESA/CSA James Webb Space Telescope. All raw and pipeline-processed data presented in this article are available via the MAST archive DOI: \url{10.17909/rwee-wr25}. The custom-reduced data products described in this article are available on Zenodo at DOI \url{10.5281/zenodo.12531991}. MAST is operated by the Association of Universities for Research in Astronomy, Inc., under NASA contract NAS 5-03127 for JWST. These observations are associated with program \#2965.

This research was supported in part through computational resources and services provided by Advanced Research Computing at the University of Michigan, Ann Arbor. 

This research has made use of the NASA Exoplanet Archive, which is operated by the California Institute of Technology, under contract with the National Aeronautics and Space Administration under the Exoplanet Exploration Program. 

JMV acknowledges support from a Royal Society - Research Ireland University Research Fellowship (URF/1/221932). BB and BJS acknowledge funding by the UK Science and Technology Facilities Council (STFC) grant no. ST/V000594/1.
BK, MJW and AMM acknowledge support from the National Science Foundation Graduate Research Fellowship Program under Grant Nos.~DGE-2241144, DGE-2240310 and DGE-1840990, respectively.

\facilities{JWST.}

\software{{\tt Astropy} \citep{2013A&A...558A..33A, 2018AJ....156..123A, 2022ApJ...935..167A}, {\tt NumPy} \citep{harris2020array}, {\tt dynesty.py}  \citep{2020MNRAS.493.3132S, sergey_koposov_2024_12537467}, {\tt corner.py} \citep{corner}, {\tt Matplotlib} \citep{Hunter:2007}, {\tt pandas} \citep{mckinney-proc-scipy-2010}, \& { the transit search code used in this manuscript will be made available upon request.}}

\bibliography{main}{}
\bibliographystyle{aasjournal}
\end{document}